\documentclass[sigconf]{acmart}

\AtBeginDocument{%
  \providecommand\BibTeX{{%
    \normalfont B\kern-0.5em{\scshape i\kern-0.25em b}\kern-0.8em\TeX}}}

\copyrightyear{2024}
\acmYear{2024}
\setcopyright{rightsretained}
\acmConference[WSDM '24]{Proceedings of the 17th ACM International Conference on Web Search and Data Mining}{March 4--8, 2024}{Merida, Mexico}
\acmBooktitle{Proceedings of the 17th ACM International Conference on Web Search and Data Mining (WSDM '24), March 4--8, 2024, Merida, Mexico}
\acmDOI{10.1145/3616855.3635736}
\acmISBN{979-8-4007-0371-3/24/03}

\begin{document}
\title{Scaling Up LLM Reviews for Google Ads Content Moderation}




\author{Wei Qiao}
\affiliation{%
  \institution{Google Ads Safety}
}

\author{Tushar Dogra}
\affiliation{%
  \institution{Google Ads Safety}
}

\author{Otilia Stretcu}
\affiliation{%
  \institution{Google Research}
}

\author{Yu-Han Lyu}
\affiliation{%
  \institution{Google Ads Safety}
}

\author{Tiantian Fang}
\affiliation{%
  \institution{Google Ads Safety}
}

\author{Dongjin Kwon}
\affiliation{%
  \institution{Google Ads Safety}
}

\author{Chun-Ta Lu}
\affiliation{%
  \institution{Google Research}
}

\author{Enming Luo}
\affiliation{%
  \institution{Google Research}
}

\author{Yuan Wang}
\affiliation{%
  \institution{Google Ads Safety}
}

\author{Chih-Chun Chia}
\affiliation{%
  \institution{Google Ads Safety}
}

\author{Ariel Fuxman}
\affiliation{%
  \institution{Google Research}
}

\author{Fangzhou Wang}
\affiliation{%
  \institution{Google Ads Safety}
}

\author{Ranjay Krishna}
\affiliation{%
  \institution{Univ. of Washington}
}

\author{Mehmet Tek}
\affiliation{%
  \institution{Google Ads Safety}
}

\renewcommand{\shortauthors}{Wei Qiao, et al.}

\begin{abstract}
Large language models (LLMs) are powerful tools for content moderation, but their inference costs and latency make them prohibitive for casual use on large datasets, such as the Google Ads repository. This study proposes a method for scaling up LLM reviews for content moderation in Google Ads. First, we use heuristics to select candidates via filtering and duplicate removal, and create clusters of ads for which we select one representative ad per cluster. We then use LLMs to review only the representative ads. Finally, we propagate the LLM decisions for the representative ads back to their clusters. This method reduces the number of reviews by more than 3 orders of magnitude while achieving a 2x recall compared to a baseline non-LLM model. The success of this approach is a strong function of the representations used in clustering and label propagation; we found that cross-modal similarity representations yield better results than uni-modal representations.
\end{abstract}

\vspace{-0.1cm}

\maketitle
\section{Outline}
We introduce an end-to-end solution for scaling up and leveraging LLMs~\cite{chen2023pali,anil2023palm} for Google ads content moderation. We will first provide the context for the LLM review problem and explain the scale of reviews in the context of compute resources. Then, we will introduce our approach to solving this problem. We will conclude with the results of our deployment in Google Ads policy enforcement platform. Finally, we discuss future improvements and extensions.

\vspace{-0.1cm}

\section{Problem and Motivation}
Our goal is to accurately detect Google Ads policy violations in all the ads traffic before any ads are eligible to enter the auction for serving. We evaluated and used this technique on image ads only, but the approach is generic and can be extended to any modality and ad format. In this paper, we use the term ``LLMs'' to include both large language models and large visual-language models. Using LLMs to moderate all image ads traffic requires significant compute resources, making it impractical. Collecting human annotated data for fine-tuning or training a small model is also expensive because of limited human review bandwidth. Therefore, we took Google's existing LLMs, and used prompt engineering / tuning to achieve a high-quality LLM for Ads content moderation, and then scaled this model to achieve maximum recall with minimal compute resources. We evaluated our approach on the "Non-Family Safe" ad content policy, which restricts any Sexually Suggestive, Sexual Merchandise, Nudity and so on, since this is one of the important policies to protect users, advertisers and publishers. 

\begin{figure}[t]
  \centering
  \includegraphics[width=\linewidth]{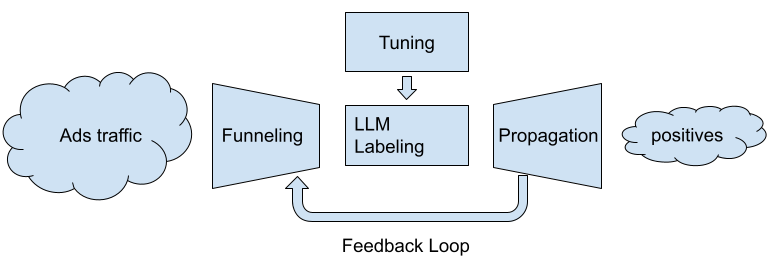}
  \caption{A diagram of our end-to-end solution for scaling up LLMs for content moderation.}
  \Description{An end-to-end solution that scales up LLM reviews for ads content moderation.}
   \vspace{-0.5cm}
\end{figure}

\vspace{-0.1cm}

\section{Method}
At a high level, our approach combines funneling, LLM labeling, label propagation, and a feedback loop. Funneling, or the review candidate selection, reduces the volume of content that needs to be processed by the LLM by using heuristic (content similarity, actor similarity, non-LLM model scores) based selection, hash based deduping, activity based filtering, and cluster-based sampling. Next, we run inference using a prompt-engineered and tuned LLM. Then, the label propagation uses a content similarity-based technique to boost the impact. Finally, a feedback loop from the final labeled images (by the LLM directly and through propagation) to the initial funneling step helps to select similar candidate images to the already labeled images in the subsequent rounds of funneling, expanding the LLM coverage across the entire image ads traffic.

\subsection{Review Candidate Selection Funneling}
We use various heuristics and signals to select potential policy-violating candidates, and then do filtering and diversified sampling to reduce the volume that needs to be processed by the LLMs.

\subsubsection{\bfseries Selecting Possible Policy Violating Candidates}
We use content and actor similarity to select an initial, larger pool of candidates. For content similarity, we leverage a graph-based label propagation technique to propagate labels from known policy-violating images as the source images (from past human or model labeled images) to similar images based on pre-trained embeddings. Two images whose distance in the embedding space is less than a threshold are considered similar. We build a similarity graph to collect the neighbors of known policy-violating content. For actor similarity, we collect candidate ad images from the accounts with policy-violating activities. 

To select candidate images with scores larger than the given thresholds, we use pre-trained non-LLM models in some cases. Using pre-trained models for candidate selection has lower precision requirements than using them for labeling.

\subsubsection{\bfseries Reducing the Pool by Deduping, Filtering, Sampling}
Google ads contains a lot of duplicate or near-duplicate content, which wastes machine resources on processing similar content. To avoid this, we first run cross-day deduping to remove images already reviewed by LLMs in the past. Then we run intra-batch deduping to only send unique images to LLMs. We also filter out inactive images and those already labeled. To perform diversified sampling, we use graph based maximal coverage sampling to sample images with diversity.

\subsection{Large Language Model Tuning and Labeling}
To adapt an LLM to a given task, one can use different strategies, such as prompt engineering~\cite{reynolds2021prompt} and parameter efficient tuning~\cite{lester2021power, hu2021lora}. Prompt engineering involves carefully designing the questions that are asked of the LLM, while parameter efficient tuning involves fine-tuning an LLM with fewer parameters on a labeled dataset to adjust its parameters to the task at hand. In our work, we took advantage of the ability of LLMs to do in-context learning~\citep{brown2020language}, and used a combination of prompt engineering and parameter efficient tuning to prepare an LLM that performs well on our policy.

To validate the model's performance on manually curated prompts, policy experts first performed prompt engineering. For example, for a Non-Family Safe policy, we might prompt the LLM with a question such as "Does the image contain sexually suggestive content?". The LLM's predictions are then parsed into a binary yes/no policy label.
Because the LLM's accuracy varies depending on the prompt, our policy experts crafted and evaluated various prompts on a small labeled dataset in order to select the best-performing prompt for our task, which was then used in combination with soft-prompt tuning~\cite{lester2021power} to create the final prompt used by our production system. During soft-prompt tuning, a small uninterpretable prompt is trained to nudge the LLM towards the correct answers on a labeled training set. This has been shown in the literature to significantly improve LLM performance~\cite{lester2021power}, and we observed the same in our experiments.

Note that prompt engineering and tuning are one-time costs, performed {\em only once per policy}. Once the prompt is constructed, it can be used for all inference runs of our system. For each candidate we want to classify with an LLM, we concatenate the prompt and the image and pass them to the LLM for labeling.

\subsection{Label Propagation and Feedback Loop}
From LLM labeled candidates of the previous stage, we propagate the label of each image to the similar images from stored images we've seen in the past traffic. We store selected LLM labeled images as known images and label incoming images if they are similar enough to be considered as near duplicates.

All labeled images, whether directly by LLMs or indirectly labeled through label propagation, are then read in the review candidate selection stage, and used as input in the initial known images for content similarity based expansion, to identify similar images as potential candidates for the next round of LLM review.

\section{Results and Discussions}
We ran our pipeline over 400 million ad images collected over the last 30 days. Through funneling, we reduced the volume to less than 0.1\%, or 400k images, which are reviewed by an LLM. After label propagation, the number of ads with positive labels doubled. This pipeline labeled roughly twice as many images as a multi-modal non-LLM model, while also surpassing its precision on the ``Non-Family Safe'' ad policy. Overall, this pipeline helped remove more than 15\% of the policy-violating impressions among image ads for this policy. 

We are expanding this technique to more ad policies and modalities, such as videos, text, and landing pages. We are also improving the quality of all pipeline stages, including funneling by exploring better heuristics, tuning better LLM prompts, and propagating similarity through higher-quality embeddings.

\section{Company Portrait}
{\bfseries Google LLC} is an AI-first multinational company focused on organizing the world’s information and making it universally accessible and useful. Google operates businesses in online advertising, search engine technology, cloud computing, and consumer electronics.

\section{Presenter Biography}
{\bfseries Wei Qiao}: Wei is a technical lead in Google Ads Content and Targeting Safety team. He is leading efforts to build the systems and workflows for efficient ads content moderation. Contact email: \href{mailto:weiqiao@google.com}{weiqiao@google.com}.

\bibliographystyle{ACM-Reference-Format}
\bibliography{reference}

\clearpage

\end{document}